\documentclass[manuscript,screen]{acmart}

\AtBeginDocument{%
  \providecommand\BibTeX{{%
    \normalfont B\kern-0.5em{\scshape i\kern-0.25em b}\kern-0.8em\TeX}}}

\setcopyright{rightsretained}
\copyrightyear{2023}
\acmYear{2023}
\acmDOI{10.13140/RG.2.2.23377.10085/1}

\acmConference[Making A Real Connection, MUM ’23]{Making A Real Connection, Pro-Social Collaborative Play in Extended Realities – Trends, Challenges and Potentials}{Dec. 03--06, 2023}{Vienna, Austria}
\acmBooktitle{Workshop "Making A Real Connection, Pro-Social Collaborative Play in Extended Realities – Trends, Challenges and Potentials" at 22nd International Conference on Mobile and Ubiquitous Multimedia (Making A Real Connection, MUM ’23), December 03--06, 2023, Vienna, Austria}

\begin{document}

\title[Techn. Challenges of Ambient Serious Games in Higher Education]{Technological Challenges of Ambient Serious Games in Higher Education}

\author{Lea C. Brandl}
\authornote{Both authors contributed equally to this research.}
\email{lea.brandl@uni-luebeck.de}
\orcid{0000-0001-6655-6763}
\author{Börge Kordts}
\authornotemark[1]
\email{b.kordts@uni-luebeck.de}
\orcid{0000-0002-4235-8399}

\author{Andreas Schrader}
\email{andreas.schrader@uni-luebeck.de}
\orcid{0000-0001-7926-0611}
\affiliation{%
  \institution{Institute of Telematics, University of Lübeck}
  \streetaddress{Ratzeburger Allee 160}
  \city{23562 Lübeck}
  \country{Germany}}

\renewcommand{\shortauthors}{Brandl and Kordts, et al.}

\begin{abstract}
Naturally, university courses should be designed to attract students, engaging them to achieve learning goals. Toward this end, the use of Serious Games has been proposed in the literature. 
To address positive effects, such as content memorability and attendance rates, we propose Ambient Serious Games as games embedded in a computer-enriched environment, which is only partially perceived mentally by players. In this paper, we describe five technological key challenges that must be overcome to seamlessly and beneficially integrate an Ambient Serious Game into teaching. These challenges, derived from a scenario, focus on the technological provision and conduct of such games based on a software platform. They include (1) the integration of physical smart learning objects in heterogeneous environments under dynamic constraints, (2) the representation of abstract subject matter using smart learning objects, (3) the guided or automatic connection of all involved components, (4) the explanation of the components, their interaction, as well as the serious game itself, and (5) feedback on the game state.
\end{abstract}




\maketitle

\section{Introduction}

The attendance rate in university courses is directly related to learning success and the associated examination performance \cite{crede2010}.
Consequently, courses should be made attractive for students. The use of activating teaching methods represents a possible approach in this regard \cite{bochmann2018}.
Roepke et al. \cite{roepke2019} investigated the use of such methods in different subject areas and found that activating methods are used less frequently than conventional approaches. However, according to Dale's cone of experience \cite{dale1969}, activating methods should have a beneficial effect on knowledge transfer.

As possible countermeasures, Roepke et al. suggest incorporating puzzles, quizzes, or the games Taboo, as well as Activity, into the teaching routine. 
Such games are categorized as Serious Games, i.e., games that do not focus on entertainment, fun, or amusement as their primary or sole purpose \cite{michael2005}. 
Learning games can have a motivating effect on students and can cause them to engage with content longer than in other teaching formats \cite{faiella2015,boeker2013,susi2007}. Backlund and Hendrix \cite{backlund2013} report mostly positive effects on learning impact, problem-solving skills, and motivation to learn. However, they also point out that in some cases the results seem to be influenced by whether the evaluation was conducted by the game's developers or others (observer bias).

Generally, the cognitive learning process should be taken into account, whereby especially the Cognitive Theory of Multimedia Learning can play a role. It explains a higher cognitive availability when using multimedia to present learning content, particularly combining visual and auditory presentation \cite{mayer2005}. 
Hence, using multimedia to show the same teaching content can allow for more efficient information processing and seems to increase memorability.

Consequently, the paradigm shift observable in other domains towards embedding technology in the ambience (known as \emph{Pervasive Computing}, cf. \cite{hansmann2003pervasive}) could also be used to design courses that take into account the aspects mentioned above. In doing so, interactive learning spaces could integrate activating teaching methods, multimodal learning and tangible learning objects designed to illustrate subject matter.
The vision of an adaptive learning space available as a learning management system has already been proposed by various authors \cite{rahimi2022,olsevicova2008,mikulecky2005,tavangarian2009}. It includes the design of pervasive environments in the university environment for the presentation and teaching of learning content. 

The paradigm of Pervasive Computing has also been applied to games resulting in \textit{Pervasive Games}, i.e., games that are not clearly limited in time, space, or social conditions \cite{montola2005}. More specialized, Ambient Games represent a subset, where the atmosphere in which players engage in various activities is essential, including non-game activities. They allow players to use information provided by the environment without forcing interaction \cite{eyles2008}. They incorporate the concept of Pervasive Computing and exploit the augmentation of reality through the use of embodied virtuality.

To address the above-mentioned aspects, we propose the conjunction of Ambient and Serious Games. Hence, we define the term \emph{Ambient Serious Game} as follows: an Ambient Serious Game is a game that has a serious goal. It is embedded in an environment enriched with computers, which is only partially mentally perceived by the players. Nevertheless, the focus is on the game as such, which includes a set of rules as well as a defined goal. This definition partly contradicts the definition of Ambient Games mentioned above, but ambient in this case describes the use of smart environment whose technical components are not mentally perceivable by the users. In the university context, Ambient Serious Games can be understood as learning games that are integrated into lectures in order to achieve positive effects.

In this paper, we broach the issue of the following research question: what are the technological key challenges that need to be overcome to seamlessly and beneficially integrate an Ambient Serious Game into teaching? Towards this end, we present a scenario of the integration of an Ambient Serious Game into teaching routine that illustrates possible benefits and serves as a basis for deriving technical issues.

\section{User Scenario}
\begin{figure}[t]
    \centering
    \includegraphics[width=\textwidth]{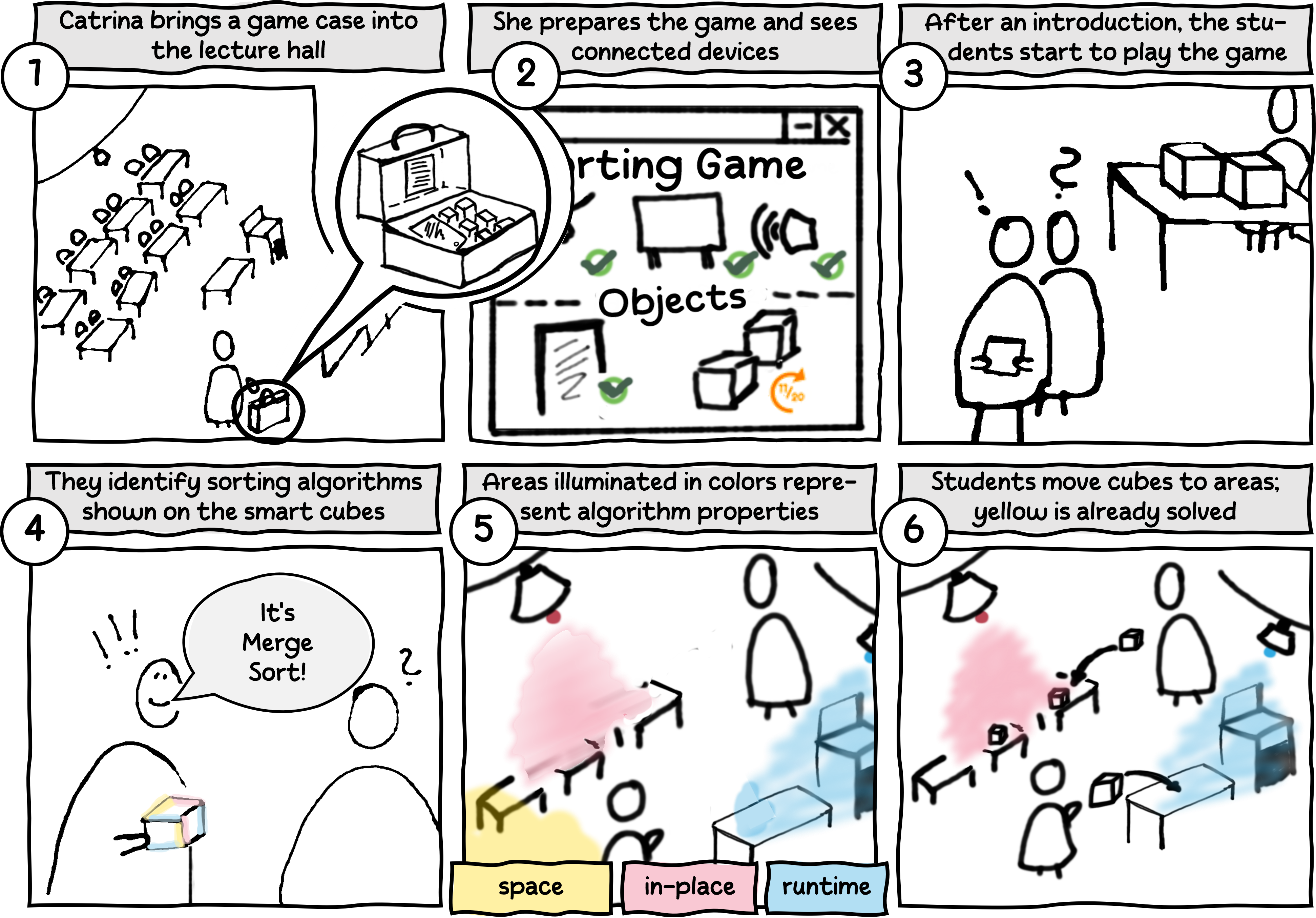}
    \caption{Illustration of the scenario. Frames 1 and 2 illustrate the set-up process. Frames 3 to 6 illustrate the students playing the game. The processes shown are repeated with other assignments.}
    \label{fig:szenario}
\end{figure}

The following scenario illustrates how a simple Ambient Serious Game could be integrated into lectures. For easier understanding, the scenario is also graphically underlined by Fig.~\ref{fig:szenario}.

Catrina Smith is a research assistant at the Institute of Computer Science. Among other lectures, she also gives a lecture on the topic of sorting algorithms. In order to make the course more attractive for the students, she plans to integrate an Ambient Serious Game in her lecture. Since she needs additional equipment for the game, she borrows a specially equipped game case from IT support and brings it with her into the lecture hall (see Fig.~\ref{fig:szenario}, Frame~1).

Once in the lecture hall, she opens the case. It contains some smart cubes and a tablet. She also finds instructions in the lid. Following the instructions, she uses her computer to open the exercise in the learning management system of the university. By clicking on the icon labelled \emph{Ambient Sorting Game}, she can now start the exercise as a pervasive multiplayer game. In the next step, the system shows that it has discovered the cubes, the tablet, the room's smart lights as well as the smart board (which is also connected to the audio system) of the lecture hall (see Frame~2). The system now displays further instructions, telling Catrina to place the cubes that were in the case on the table. Both the cubes and the smart lights flash briefly, from which Catrina concludes that everything is connected and set up. First students have already taken their seats in the auditorium and shortly afterward, Catrina begins her lecture.

In the first half hour, she explains various sorting algorithms. After that, Catrina starts the game and introduces it to the students. She also explains the learning objectives. Instructions for the players are displayed on the smart board. Background music can be heard over the speakers, its calm nature suggesting that there is still plenty of time to process. The students work with the smart cubes that now display sorting algorithms in action (see Frame~3), which the students have to identify (see Frame~4). Following this task, the students are urged to assign them to color-coded playing areas, which depict properties of some of the sorting algorithms. The game tells the students that these color fields are projected into the auditorium by the lighting. The students look around and notice three areas in the room illuminated in different colors: red, blue, and yellow (see Frame~5). The tablet states that red is assigned to the property "in-place", blue to "optimal runtime" and yellow "optimal space complexity".  
Also, the smart cubes begin to glow in the different colors. The students assume that they can now move the cubes to the colored area. 
To place the cubes, the students need to walk back and forth between the areas. Once they solved "optimal space complexity" by correctly putting all cubes in the yellow area, the yellow light turns off (see Frame~6). In the meantime, the background music has become more tense, indicating that there is not much time left. Finally, the students are just about to place the last cube. A rewarding jingle can be heard upon placing the last cube, and the current score as well as the remaining time can be seen on the smart board. 

The students can now see that they have achieved 23,847 points. The leaderboard displayed on the smart board states that they have scored second-best all-time. Following the game, Catrina discusses the decisions of the students and the results to clarify misunderstandings. Afterward, the students are more attentive and can concentrate well on the new material following up.

\section{What is challenging?}

Some obstacles to the widespread use of games in the classroom have already been reported. For example, Backlund and Hendrix \cite{backlund2013} report barriers, including acceptance issues and technological restrictions, which are also reflected in our scenario. However, their review focuses on educational games and does not explicitly address Ambient Serious Games. Thus, further technological challenges need to be addressed.

% Challenge I
First of all, to allow students to interact with the environment in the course of a game, classrooms need to be transformed into interactive learning spaces. So far, rooms used for education are typically not equipped with various smart objects that are ready to be used. Instead, most of the required equipment needs to be brought into the room. The scenario also illustrates that even when fully equipped classrooms are present, additional (tangible) learning objects can be required. So, even in the lecture hall with smart lights and a smart board, Catrina needs to bring a game case with additional equipment to her lecture. Furthermore, devices as well as persons involved may enter or leave the space dynamically. Hence, a platform for an Ambient Serious Game needs to cope with heterogeneously equipped learning spaces, further components brought by the lecturers, as well as dynamically changing situations.

% Challenge II
Secondly, tangible as well as abstract lecture content needs to be covered by the game. While some subject matter can be directly presented using physical representatives or tangible objects, a large proportion covers abstract topics or information that is not directly associated with a physical form or representative objects. In the presented scenario, the smart cubes are used to visualize sorting algorithms which are abstract by nature. Thus, at least some of the used smart learning objects need to be able to display abstract subject matter. Ideally, both tangible representative smart objects and generic smart objects with display capabilities are combined.

% Challenge III
Next, the provision of an Ambient Serious Game should require as little effort as possible to gain acceptance. Ideally, setting up the equipment in the interactive learning space and launching the game are part of a facile setup routine. Catrina just needs to bring the game case, distribute the smart cubes as well as the tablet, and start the application via the learning management system. The system in the scenario discovers and connects the components by itself. Thus, a platform for an Ambient Serious Game should automatically discover smart objects as well as further involved devices and be able to connect all involved components, including the game itself. Ideally, the platform is able to make a proposal for possible and sensible connections between the involved components as well as the game. Though, decision-making authority should remain with the lecturers. They should be able to control or manipulate the connection process and overrule proposals, or part of a proposal, made by the platform. 

% Challenge IV
Also, students as well as lecturers should know how to play the game. Baalsrud et al. \cite{baalsrud2021} observed that it is important that lecturers are well-informed about the game itself and the interaction technologies. However, the resulting interplay of devices and the game in the interactive learning space can be highly dynamic. Also, one game can be played using various different (and interchangeable) ensembles of smart learning objects and vice versa. It should be noted that the actual game (experience) arises out of said interplay of all involved components. However, the automatic connection of involved components and the dynamic situation may obfuscate the handling of the game. In our scenario, Catrina as well as the students are instructed by the system at any time. In addition to the instructions displayed on the smart board, the smart objects also provide information on correct use. 
Thus, the platform should instruct the students or the lecturers, at best both, on how it is controlled using adequate modalities. Ideally, each component has a knowledge of its internal state, the connection to other devices as well as the consequences of possible actions and further, is able to explain itself based on this knowledge (self-reflection, cf. \cite{burmeister2017}).

% Challenge V
Finally, involved players as well as lecturers should be informed about the state of the game. Hence, the game should provide continuous and multimodal feedback to both parties (cf. \cite{caserman2020}). The scenario shows multiple situations in which feedback is provided. While the smart board displays the game statistics all the time, smart lights indicate areas for placing the smart cubes and flashing lights on the smart cubes depict an established connection. Thus, the platform needs to be able to locate devices and their relative positions to provide useful feedback regarding progress and correct or incorrect decisions by students. Ideally, feedback is provided at the right time at the right position (i.e., close to interaction). Furthermore, feedback mechanisms may be required to be explained using the aforementioned instructions for players to fully understand game operations. 

\section{Discussion}

While introducing Ambient Serious Games into university teaching routines can potentially address aspects such as the memorability of subject matter or attendance rates, a software platform for the provision and conduct of such games would need to overcome several challenges. We have identified five key challenges regarding the interplay of smart objects in heterogeneous environments, the physical representation of abstract content, the connection of involved components, explanations for all components, and feedback on the game state.

The described challenges essentially refer to technical aspects in the provision and conduct of Ambient Serious Games in teaching routine. They do not yet address the concrete design of such games and, in particular, the integration of teaching content. The introduction to the game and debriefing afterward go beyond self-explainability but can extend and support it. In our scenario, the lecturer handles both the introduction and debriefing. Visionarily, the software platform could take on these roles.

Additionally, the implementation of Ambient Serious Games requires basic technical equipment in the lecture hall, depending on the respective game, the number of students, and possibly other circumstances. It can be expected that learning spaces will become increasingly enriched with technology, and the involved technology will become more advanced. This would allow games to become more immersive, and game elements to be more sophisticated. For instance, a smart floor could be used to illuminate clearly delimited areas for assignment tasks similar to the one presented in our scenario.

More complex games and scenarios are also conceivable. For example, the described game could be designed collaboratively for groups. The colored spaces in the room could be unlocked by a second group answering questions. The students could also assign properties of the algorithms to these areas. This would require strategic coordination and mutual support between the groups in order to advance faster.

While we expect that addressing the identified challenges ensures a certain acceptance among lecturers, it should be noted that a minimum degree of openness to new methods and technologies is required. Particularly, the interweaving of subject matter, game design, and arrangement of the smart learning spaces plays an important role, which should be planned carefully. Furthermore, individual learning styles, as well as player types, should be considered.

To substantiate the identified challenges with respect to stakeholders' perspectives, we plan to conduct studies regarding the integration into teaching routines and the design of concrete games. We further plan to examine acceptance, particularly potential barriers and enablers, as well as evidence regarding outcomes in terms of learning success. In particular, distinct games need to be evaluated to prove their effectiveness. In doing so, we plan to address further, more broadly framed research questions regarding the usefulness of the design concept of Ambient Serious Games and smart learning spaces.

\bibliographystyle{ACM-Reference-Format}
\bibliography{sample-manuscript}

\end{document}